\documentclass[10pt]{article}
\usepackage[centertags]{amsmath}
\usepackage{amssymb}

\newcommand{\rmd}{{\rm d}}
\newcommand{\rme}{{\rm e}}
\newcommand{\rmi}{{\rm i}}

\newcommand{\la}{\langle}
\newcommand{\ra}{\rangle}
\newcommand{\lla}{\left\langle}
\newcommand{\rra}{\right\rangle}

\newcommand{\cN}{{\cal N}}
\newcommand{\cM}{{\cal M}}

\newcommand{\ts}{\textstyle}

\begin{document}

\title{Integrals of monomials over the orthogonal group}

\author{T. Gorin\footnote{Thomas.Gorin@physik.uni-freiburg.de}\\
{\it\small Centro de Ciencias F\'\i sicas, University of Mexico (UNAM), 
           Avenida Universidad s/n,}\\
{\it\small C.P. 62210 Cuernavaca, Morelos, M\' exico}\\[2mm]
{\it\small Theoretische Quantendynamik, Fakult\" at f\" ur Physik, 
           Universit\" at Freiburg, Hermann-Herder-Str. 3,}\\
{\it\small D-79104 Freiburg, Germany}
}

\date{}

\maketitle

\begin{abstract}
A recursion formula is derived which allows to evaluate invariant integrals 
over the orthogonal group $O(N)$, where the integrand is an arbitrary finite 
monomial in the matrix elements of the group. The value of such an 
integral is expressible as a finite sum of partial fractions in $N$. The 
recursion formula largely extends presently available integration formulas for 
the orthogonal group. \\

\noindent
PACS: 02.20.-a, 
      05.40.-a, 
      05.45.Mt  

\end{abstract}

\section{Introduction \label{I}}

Integrals over the classical compact groups$^{1,2}$ are of interest in various 
fields, such as harmonic analysis$^3$ or random matrix theory$^4$. 
In these applications the integrand is often a polynomial in the matrix 
elements of the group itself, ({\it i.e.} of the true matrix
representation of the group). Thus we have to integrate an arbitrary monomial
of these matrix elements. Closed formulas are available only for very special
cases$^{3,5-7}$ and even a new method by Prosen {\it et al.}$^8$ using 
computer algebras is practically limitted to low degrees. Though note that, 
for arbitrary monomials there is strong evidence, that the results are exact 
at least up to the next leading order in $N^{-1}$ with respect to the 
approximation of the group integral by independent Gaussian distributed 
matrix elements.$^9$

In the present paper we shall address the case of the orthogonal group
$O(N)$. First we rederive the well known one-vector formula.$^{4,6}$ In
this context, the terms ``$R$-vector formula'' or ``$R$-vector integral'' 
refer to the case where the monomial in question contains only powers of 
matrix elements from $R$ rows or $R$ columns respectively. Next we derive 
a recursion formula that relates an $R$-vector integral to a linear 
combination of $(R$$-$$1)$-vector integrals. This is the central result of 
the present paper. Together with the one-vector formula, it allows to 
calculate any integral over a monomial of finite degree in a finite number 
of steps. This result is then used, to obtain a closed expression for 
general two-vector integrals that is much simpler than the one known 
before.$^6$ Besides, the older formula contains mistakes which (to the 
best of my knowledge) had never been corrected in the literature.  

The paper is organized as follows: 
In Sec.~\ref{G} we describe the current approach to the problem. In
addition, we introduce some compact non-standard notations, which help to 
keep the mathematical expressions manageable. Then the one-vector result of 
Ullah$^6$ is rederived, as it is the base for the recursion formula developed 
later on. In passing we obtain an equally simple formula for the 
corresponding one-vector integral over the unitary group. 
In Sec.~\ref{OR} we derive the general recursion formula. 
In Sec.~\ref{A} some applications are presented. As an immediate 
consequence, we obtain a closed expression for the two-vector integral, 
which is then compared to the corrected old result.$^6$ We also illustrate 
the use of our general formula for $R>2$, calculating a particular 
three-vector integral. 
Sec.~\ref{S} contains the conclusions.

\section{\label{G} General considerations}

To be specific, let us consider the orthogonal matrix $w\in O(N)$ as a point 
in Euclidean $N^2$-dimensional space. Then we are interested in integrals of 
monomials in the coordinates of $w$. These are denoted by:
\begin{equation}
\la M\ra = \int\rmd\sigma(w) \; \prod_{i,\xi=1}^{N,R} w_{i\xi}^{M_{i\xi}} \; .
\end{equation}
Here $\sigma$ is the normalized Haar measure$^{10}$ of $O(N)$, {\it i.e.} 
$\int\rmd\sigma(w) = 1$, and $M$ is a $N$$\times$$R$ matrix of non-negative 
integers, with $R\le N$. $M$ is called the power matrix. In the recursion 
formula to be developed, $R$ is used as the recursion parameter. Hence it is 
important, that $R$, the number of columns of $M$, is as small as possible. 

The integral over the orthogonal group is invariant under any permutation of
columns or rows of the integration variable $w\in O(N)$. It is also invariant 
under taking the transpose. Therefore it is sufficient to consider such
monomials which contain matrix elements from the first $R\le N$ columns
of $w$ only. According to Ullah$^6$ one may then write:
\begin{equation}
\la M\ra  = \frac{\cN(M)}{\cN(o)} \; , \quad
\cN(M) = \int\prod_{\xi=1}^R \left\{ \rmd\Omega(\vec w_\xi) \prod_{i=1}^N
w_{i\xi}^{M_{i\xi}} \right\} \; 
\prod_{\mu <\nu} \delta(\la\vec w_\mu|\vec w_\nu\ra) \; ,
\label{G_avM}\end{equation}
where $o$ is a $N$$\times$$R$ matrix of zeros. The integration region is 
the Cartesian product of $R$ unit spheres with constant measures 
$\rmd\Omega(\vec w_\xi)$, and $\{\vec w_1,\ldots,\vec w_R\}$ are the 
corresponding unit vectors. The orthogonality of the unit vectors is 
implemented with the help of appropriately chosen $\delta$-functions.
$\la\vec w_\mu|\vec w_\nu\ra$ denotes the scalar product between the two
vectors: $\vec w_\mu$ and $\vec w_\nu$.

\subsection{Compact notations for certain products of multinomials}

In the calculations to follow, we will frequently deal with certain products
of binomial and multinomial coefficients. For better legibility we use two
special notations: In what follows, $\vec x$ and $\vec y$ are 
$N$-dimensional real vectors, and $\vec m$ and $\vec n$ are $N$-dimensional 
vector-indices of non-negative integers. There are two typical cases in which 
products of multinomials appear: 

\begin{itemize}
\item[1)]{Consider the expression $E= \prod_{i=1}^N (x_i+y_i)^{n_i}$. Its
expansion gives:
\begin{equation}
E = \sum_{\vec k} \; \prod_{i=1}^N {n_i \choose k_i} \; x_i^{k_i} \;
y_i^{n_i-k_i} \; ,
\end{equation}
where the sum runs over all $\vec k$ for which 
$\forall i : k_i \le n_i$. In this case, the product of binomials is denoted
by the following symbol:
\begin{equation}
\prod_{i=1}^N {n_i \choose k_i} = {\vec n \choose \vec k} \; .
\label{G_shortbin}\end{equation}
}
\item[2)]{In other occasions, we encounter expressions of the type:
$E'=\prod_{i=1}^N\la\vec\tau|\vec w_i'\ra^{n_i}$, where 
$\la\vec\tau|\vec w_i'\ra$ is the scalar product of the two 
$(R$$-$$1)$-dimensional vectors $\vec\tau$ and $\vec w_i'$. In this case 
the expansion reads:
\begin{equation}
E'= \sum_K \left\{ \prod_{i=1}^N (n_i|K_{i1},\ldots,K_{i,R-1})
\right\} \; \prod_{\xi=1}^{R-1} \tau_\xi^{\bar k_\xi} \prod_{i=1}^N
w_{i\xi}^{K_{i\xi}} \; .
\label{G_Eprime}\end{equation}
Here we have to use the $N$$\times$$(R$$-$$1)$ matrix $K$ as an index. The
elements of $K$ are non-negative integers. $\vec k_\xi$ is the $\xi$th 
column vector of $K$, and $\bar k_\xi$ is the sum of its components:
$\bar k_\xi = \sum_{i=1}^N K_{i\xi}$. The sum in Eq.~(\ref{G_Eprime})
runs over all $K$ for which $\forall i : \sum_{\xi=1}^{R-1} K_{i\xi} = n_i$. 
In this case, we use the following notation:
\begin{equation}
\prod_{i=1}^N (n_i|K_{i1},\ldots,K_{i,R-1}) = (\vec n\, | K) \; .
\label{G_shortmul}\end{equation}
}
\end{itemize}

\subsection{\label{O1}The one-vector integral}

In the one-vector case, $R=1$, there are no orthogonality relations to
respect. The power matrix $M$ consists of one single column vector, here
denoted by $\vec m$. According to Eq.~(\ref{G_avM}) we may write:
\begin{equation}
\la\vec m\ra = \frac{\cN(\vec m)}{\cN(\vec o)} \; , \quad
\cN(\vec m) = \int\rmd\Omega(\vec w) \prod_{i=1}^N w_i^{m_i} \; ,
\end{equation}
where $\vec o$ is a $N$-dimensional vector of zeros. Following the original 
calculation of Ullah,$^{4,6}$ we integrate over the full vector space 
$\mathbb{R}^N$ and implement the normalization of the column vector with the 
help of a $\delta$-function. This introduces the integration constant 
$c_1(N)$:
\begin{equation}
\cN(\vec m) = c_1(N)^{-1} \left\{\prod_{i=1}^N\int_{-\infty}^\infty\rmd x_i\;
x_i^{m_i} \; \right\} \; \delta\left(\|\vec x\|^2 - 1\right) \; .
\label{O1_2}\end{equation}
The next step is to remove the $\delta$-function. Setting $x_i = u_i/\sqrt{r}$, 
we obtain:
\begin{equation}
\cN(\vec m) \; r^{(N+\bar m)/2-1} = c_1(N)^{-1} \left\{
\prod_{i=1}^N \int_{-\infty}^\infty\rmd u_i \; u_i^{m_i} \; \right\} \;
\delta\left(\|\vec u\|^2 - r\right) \; ,
\end{equation}
where $\bar m =\sum_{i=1}^N m_i$. The $\delta$-function can now be removed by
multiplying the {\it l.h.s.} and the {\it r.h.s.} with $\rme^{-r}$ and 
integrating on $r$ from $0$ to $\infty$:
\begin{equation}
\cN(\vec m) \; \Gamma\left(\ts\frac{N+\bar m}{2}\right) \; c_1(N) =
\prod_{i=1}^N \int_{-\infty}^\infty\rmd u_i \; u_i^{m_i} \; \rme^{-u_i^2} =
\prod_{i=1}^N \Gamma\left(\ts\frac{1+m_i}{2}\right) \; .
\label{O1_Nvm}\end{equation}
Solving this equation for $\cN(\vec m)$, the ratio $\cN(\vec m)/\cN(\vec o)$
can be calculated, which leads to the desired result:
\begin{equation}
\la\vec m\ra = \left(\ts\frac{N}{2}\right)_{\bar m/2}^{-1} \; 
\prod_{i=1}^N \left(\ts\frac{1}{2}\right)_{m_i/2} \; .
\label{O1_res}\end{equation}
Here it is convenient to use the Pochhammer symbol
$(z)_n= \Gamma(z+n)/\Gamma(z)$.$^{11}$  Note that Eq.~(\ref{O1_Nvm}) implies, 
that the integral $\la\vec m\ra$ vanishes if at least one component of 
$\vec m$ is odd.

\subsection{The one-vector integral over the unitary group}

It is natural to consider also integrals over the unitary group $U(N)$. This
is in general much more complicated because usually the monomials to integrate 
contain powers of the matrix elements and their complex conjugated 
counterparts. However in the one-vector case, the integral over the unitary 
group can be mapped on a corresponding integral over the orthogonal group, 
which leads again to a simple result (it seems
that such a formula has never been published elsewhere). If more vectors are 
involved, $R>1$, the orthogonality conditions destroy this simple 
correspondence.

To obtain the desired expression for one-vector integrals, it is 
convenient to consider monomials in the real and imaginary parts of the complex
unit vector $\vec w$. They can be identified with the coordinates in a 
$2N$-dimensional Euclidean space, where the Haar measure reduces to the 
constant measure $\Omega_2$ on the unit hypersphere. Denoting the one-vector 
integral of an arbitrary monomial by 
$\la\vec m\! :\!\vec n\ra = \prod_{i=1}^N x_i^{m_i} y_i^{n_i}$, where
$w_i = x_i+\rmi\, y_i$, we may write:
\begin{equation}
\la\vec m\! :\!\vec n\ra = \frac{\cM(\vec m,\vec n)}{\cM(\vec o,\vec o)} \; ,
\quad \cM(\vec m,\vec n) = \int\rmd\Omega_2(\vec w) \; 
\prod_{i=1}^N x_i^{m_i} y_i^{n_i} \; , \quad w_i = x_i+\rmi\, y_i \; .
\label{gU1-1}\end{equation}
Note the different notations: $\la\vec m\! :\!\vec n\ra$ stands for the 
one-vector integral over the unitary group, while $\la\vec m,\vec n\ra$ is
used in Sec.~\ref{A} for the two-vector integral over the orthogonal group.

Equation~(\ref{gU1-1}) shows, that we may express $\la\vec m\! :\!\vec n\ra$ 
as a one-vector integral over the orthogonal group $O(2N)$: 
$\la\vec m\! :\!\vec n\ra = \la \vec p\ra$, where $\vec p$ is the 
$2N$-dimensional concatenation of $\vec m$ and $\vec n$. Then we may apply
Eq.~(\ref{O1_res}). This leads to:
\begin{equation}
\la\vec m\!:\!\vec n\ra = (N)_{(\bar m+\bar n)/2}^{-1} \prod_{i=1}^N
\left(\ts\frac{1}{2}\right)_{m_i/2} \left(\ts\frac{1}{2}\right)_{n_i/2} \; .
\end{equation}
Again the integral $\la\vec m\!:\!\vec n\ra$ vanishes, if at least one 
component of $\vec m$ or $\vec n$ is odd. \\

\section{\label{OR}The recursion formula}

The desired recursion formula shall express an arbitrary
integral $\la M\ra$, where $M$ 
is a power matrix with $R$ columns, as a linear combination of simpler 
integrals $\la M'\ra$, where $M'$ has only $R-1$ columns. Starting from 
Eq.~(\ref{G_avM}) one may attack this problem head on, and separate the 
integration on the last unit vector $\vec w_R$ from the remaining integral:
\begin{equation}
\cN(M) = \int\left\{\prod_{\xi=1}^{R-1} \rmd\Omega(\vec w_\xi) \;
\prod_{i=1}^N w_{i\xi}^{M_{i\xi}}\right\} \; \left\{
\prod_{\mu<\nu}^{R-1} \delta\left(\la\vec w_\mu|\vec w_\nu\ra\right)\right\}
\; J(\vec w_1,\ldots,\vec w_{R-1} ; \vec m_R) \; .
\label{OR_nmp}\end{equation}
Here $\vec m_R$ is the last column vector of the power matrix $M$, and
\begin{equation}
J(\vec w_1,\ldots,\vec w_{R-1} ; \vec m_R) = \int\rmd\Omega(\vec w)
\left\{\prod_{i=1}^R w_i^{M_{iR}}\right\}\;
\prod_{\xi=1}^{R-1} \delta\left(\la\vec w_\xi|\vec w\ra\right) \; .
\label{OR_la01}\end{equation}
As shown below, the value of this integral can be expressed as a linear 
combination of monomials in the integration variables 
$\{w_{i\xi}\, |\, \xi \le R-1\}$. If this is inserted back into 
Eq.~(\ref{OR_nmp}), it obviously leads to the desired recursion 
formula. \\

To evaluate the integral (\ref{OR_la01}), the integration over the unit 
sphere is replaced by an integration over the full space $\mathbb{R}^N$, 
implementing the normalization condition with the help of a
$\delta$-function. This introduces again the normalization constant 
$c_1(N)$ [cf. Eq.~(\ref{O1_2})]. The $\delta$-function is then removed 
again, using 
the same trick as in Sec.~\ref{O1}. After that, the remaining 
$\delta$-functions (responsible for the orthogonality relations) are replaced 
by their respective Fourier representations. In this way one obtains:
\begin{align}
&J(\vec w_1,\ldots,\vec w_{R-1} ; \vec m_R) =
c_1(N)^{-1} \left\{\prod_{i=1}^N \int\rmd x_i \; x_i^{M_{iR}}\right\} \;
\delta\left(\|\vec x\|^2 - 1\right) \;
\prod_{\xi=1}^{R-1} \delta\left(\la\vec w_\xi|\vec x\ra \right)
\notag\\
&\qquad\qquad\qquad
 = c_1(N)^{-1}\; \Gamma\left(\ts\frac{N-R+\bar m_R +1}{2}\right)^{-1}
\left\{\prod_{i=1}^N \int\rmd x_i \; x_i^{M_{iR}}\; \rme^{-x_i^2}
\right\} \; \prod_{\xi=1}^{R-1}
\delta\left(\la\vec w_\xi|\vec x\ra\right) \notag\\
&\qquad\qquad\qquad
 = c_1(N)^{-1}\; \Gamma\left(\ts\frac{N-R+\bar m_R +1}{2}\right)^{-1}
\int\frac{\rmd^{R-1}\vec\tau}{\pi^{R-1}}
\prod_{i=1}^N \int\rmd x_i \; x_i^{M_{iR}}\;
\rme^{-x_i^2+2\rmi\la\vec\tau|\vec w_i'\ra x_i} \; ,
\end{align}
where $\vec w_i'$ stands for the the row-vector $(w_{i1},\ldots,w_{i,R-1})^T$.
The integrals on $x_i$ are easily evaluated, leading to:
\begin{equation}
\begin{split}
J(\ldots) \; \Gamma\left(\ts \frac{N-R+\bar m_R +1}{2}\right) \; c_1(N) &=
\int\frac{\rmd^{R-1}\vec\tau}{\pi^{R-1}} \;
\rme^{-\sum_{i=1}^N \la\vec\tau|\vec w_i'\ra} \\
&\quad\times
\prod_{i=1}^N \left[\; \sum_{\kappa_i=0}^{M_{iR}} \!^{\kappa_i :\text{ even}} \;
{M_{iR}\choose \kappa_i} \;
\left(\rmi\la\vec\tau|\vec w_i'\ra\right)^{M_{iR}-\kappa_i} \;
\Gamma\left(\ts\frac{1+\kappa_i}{2}\right) \right] \; .
\end{split}
\label{OR_la0}\end{equation}
Expanding the $N$-fold product into a sum over the vector-index
$\vec\kappa = (\kappa_1,\ldots,\kappa_N)$, we obtain for the {\it l.h.s.}:
\begin{equation}
{\it l.h.s.} = \sum_{\vec\kappa} {\vec m_R \choose \vec\kappa} \;
\rmi^{\bar m_R-\bar\kappa} \; \left\{ \prod_{i=1}^N 
\Gamma\left(\ts\frac{1+\kappa_i}{2}\right)\right\} 
\int\frac{\rmd^{R-1}\vec\tau}{\pi^{R-1}} \; \left\{
\prod_{i=1}^N \la\vec\tau|\vec w_i'\ra^{M_{iR}-\kappa_i}\right\} \;
\rme^{-\la\vec\tau|A\vec\tau\ra} \; .
\label{OR_la1}\end{equation}
Note that the sum runs only over such $\vec\kappa$, for which all components 
are even, and that for the product of binomials the abbreviation from
Eq.~(\ref{G_shortbin}) is used. A bar over vector quantities such as 
$\bar m_R$ and $\bar\kappa$ denotes the sum of all their components. The
quadratic matrix $A$, with elements $A_{\mu\nu} = \la\vec w_\mu|\vec w_\nu\ra$,
has dimension $R-1$.

Now, the key observation is the following: The orthogonality conditions 
implemented in the form of $\delta$-functions in Eq.~(\ref{OR_nmp}) select 
from the total integration region only a sub-manifold. There it holds
that $\la\vec w_\mu|\vec w_\nu\ra = \delta_{\mu\nu}$, so that the matrix $A$ 
may be replaced by the unit matrix. Then it is possible to integrate the
$\vec\tau$-integral. The expansion of the product of scalar products leads to:
\begin{equation}
\begin{split}
{\it l.h.s.} &= \sum_{\vec\kappa} {\vec m_R \choose \vec\kappa} \;
\rmi^{\bar m_R-\bar\kappa} \left\{ \prod_{i=1}^N
\Gamma\left(\ts\frac{1+\kappa_i}{2}\right)\right\} \\
&\qquad\times \sum_K (\vec m_R-\vec\kappa\, | K) \;
\left\{\prod_{\xi=1}^{R-1}\int\frac{\rmd\tau_\xi}{\pi}\;
\tau_\xi^{\bar k_\xi} \; \rme^{-\tau_\xi^2} \;
\prod_{i=1}^N w_{i\xi}^{K_{i\xi}} \right\} \; ,
\end{split}
\label{OR_la2}\end{equation}
where $K$ is a matrix index with $R-1$ columns, as introduced in 
Eq.~(\ref{G_shortmul}) together with the abbreviation for the product of
multinomials. The $\bar k_\xi$'s are the sums over the components of the 
column vectors of $K$. The remaining integrals are easily evaluated, and one 
obtains:
\begin{equation}
\begin{split}
J(\ldots) &= \frac{\pi^{1-R}}
{c_1(N)\; \ts\Gamma\left(\frac{N-R+\bar m_R +1}{2}\right)} \;
\sum_{\vec\kappa}
{\vec m_R \choose \vec\kappa} \; (-1)^{(\bar m_R-\bar\kappa)/2} \;
\left\{ \prod_{i=1}^N \Gamma\left(\ts\frac{1+\kappa_i}{2}\right)\right\} \\
&\qquad\times \sum_K (\vec m_R-\vec\kappa\, | K) \;
\prod_{\xi=1}^{R-1} \Gamma\left(\ts\frac{1+\bar k_\xi}{2}\right) \;
\prod_{i=1}^N w_{i\xi}^{K_{i\xi}} \; .
\end{split}
\label{OR_la3}\end{equation}
Note that, as a consequence of the $\tau_\xi$-integrals, the sum in the second
line runs over such $K$ only, for which all $\bar k_1,\ldots\bar k_{R-1}$ are 
even. Inserting this expression into the initial Eq.~(\ref{OR_nmp})
we obtain: 
\begin{equation}
\begin{split}
\cN(M) &= \frac{\pi^{1-R}}
{c_1(N)\; \Gamma\left(\frac{N-R+\bar m_R +1}{2}\right)} \;
\sum_{\vec\kappa} {\vec m_R \choose \vec\kappa} \; 
(-1)^{(\bar m_R-\bar\kappa)/2} \; \left\{ \prod_{i=1}^N
\Gamma\left(\ts\frac{1+\kappa_i}{2}\right)\right\} \\
&\qquad\times \sum_K (\vec m_R-\vec\kappa\, | K) \; \left\{
\prod_{\xi=1}^{R-1} \Gamma\left(\ts\frac{1+\bar k_\xi}{2}\right) \right\} \;
\left\{\prod_{\xi=1}^{R-1} \int\rmd\Omega(\vec w_\xi) \;
\prod_{i=1}^N w_{i\xi}^{M_{i\xi}+K_{i\xi}}\right\} \\
&\qquad\times \left\{ \prod_{\mu<\nu}^{R-1}
\delta\left(\la\vec w_\mu|\vec w_\nu\ra \right) \right\} \; .
\end{split}
\end{equation}
The integral over the normalized vectors $\vec w_1,\ldots,\vec w_{R-1}$ can be 
identified with $\cN(M^{(R-1)}+K)$ which is a $(R$$-$$1)$-vector integral. In 
this way, we obtain a recursion formula for $\cN(M)$. For the normalization 
constant, we find:
\begin{equation}
\cN(o) = \frac{\pi^{1-R}}
{c_1(N)\; \ts\Gamma\left(\frac{N-R+1}{2}\right)} 
\left\{ \prod_{i=1}^N \Gamma\left(\ts\frac{1}{2}\right)\right\}
\left\{ \prod_{\xi=1}^{R-1} \Gamma\left(\ts\frac{1}{2}\right)\right\}
\cN(o^{(R-1)}) \; .
\end{equation}
Thus we obtain for the $R$-vector integral $\la M\ra$, defined in 
Eq.~(\ref{G_avM}):
\begin{equation}
\begin{split}
\la M\ra &= \left(\ts \frac{N-R+1}{2}\right)_{\bar m_R/2}^{-1}
\sum_{\vec\kappa} {\vec m_R \choose \vec\kappa} \;
(-1)^{(\bar m_R-\bar\kappa)/2} \; \left\{ \prod_{i=1}^N
\left(\ts\frac{1}{2}\right)_{\kappa_i/2} \right\} \\
&\qquad\times \sum_K (\vec m_R-\vec\kappa\, | K) \;
\left\{ \prod_{\xi=1}^{R-1} \left(\ts\frac{1}{2}\right)_{\bar k_\xi/2} 
\right\} \; \la M^{(R-1)}+ K\ra \; .
\end{split}
\label{OR_res}\end{equation}
This is the desired recursion formula and the main result of the present paper.
As mentioned before it is understood, that the first sum runs over such 
$\vec\kappa$ only for which all components are even, while the second runs over
such $K$ only for which all $\bar k_\xi = \sum_{i=1}^N K_{i\xi}$ are even.
Furthermore 
$\bar m_R = \sum_{i=1}^N M_{iR}\, ,\; \bar\kappa = \sum_{i=1}^N \kappa_i$, and
$M^{(R-1)}$ stands for the matrix consisting of the first $R$$-$$1$ columns of 
the matrix $M$. \\

In principle Eq.~(\ref{OR_res}) allows to evaluate integrals
of arbitrary monomials of finite degree. The result is always expressible as 
a rational function of the dimension $N$, because the repeated expansion of
Eq.~(\ref{OR_res}) leads to nested sums of partial fractions in $N$. 
In this context it is useful to note, that only the prefactor of the 
{\it r.h.s.} depends explicitly on $N$. 
The formula~(\ref{OR_res}) can be used conveniently 
if either $R$ or the degree of the monomial to be integrated are small. 
Otherwise Eq.~(\ref{OR_res}) may lead to very lengthy expressions. 
However, such expressions should still be manageable with an appropriate 
computer algebra system. This would allow for further systematic studies of 
this class of integrals. \\

In contrast to what one might expect, the integral $\la M\ra$ does not 
necessarily vanish if the power matrix $M$ has odd elements. It rather holds
the following: If any sum over a row or column of $M$ is odd, then
$\la M\ra = 0$. Though this is in fact well known,$^{12}$ it is 
still instructive to see that it follows almost immediately from the 
recursion relation (\ref{OR_res}). 

To this end, permute columns and rows, and take the transpose if necessary, to
transform $M$ in such a way that its last column contains the row or column 
whose sum of components is odd. Then apply Eq.~(\ref{OR_res}):
The sum over $\vec\kappa$ is restricted to such $\vec\kappa$ which 
have only even components. Hence $\bar\kappa$ is even. As $\bar m_R$ is odd, 
and $\sum_{\xi=1}^{R-1} \bar k_\xi = \bar m_R -\bar\kappa$, 
at least one of the sums $\bar k_1,\ldots,\bar k_{R-1}$ must be odd as well. 
Such a term vanishes, because of what is said below Eq.~(\ref{OR_la3}).
This implies that all terms of the sum over $K$ vanish likewise, so that
the complete integral gives zero.

\section{\label{A} Applications}

In random matrix theory, many matrix ensembles are based on the concept of 
orthogonal invariance. Physically this corresponds to a situation where the 
Hamiltonian for a spin-less quantum particle possesses an anti-unitary 
symmetry, {\it e.g.} time reversal invariance. The Gaussian and circular 
orthogonal ensembles$^{4,13,14}$ are well known examples based on this 
concept. Other examples are the Poisson orthogonal ensemble,$^{15}$ or the 
recently introduced matrix ensembles for semi-separable systems.$^8$ In those
cases where the orthogonal invariance applies directly to the Hamiltonian, the 
statistical properties of the eigenvectors are uniquely determined by the 
orthogonal group and its invariant measure. Therefore any correlators
between the eigenvectors can be expressed and calculated in terms of
$R$-vector integrals.\\

In what follows, we first apply our integration formula~(\ref{OR_res}) to the
two-vector case. In this way we obtain a closed expression for arbitrary
two-vector integrals. Then we compare this result with the corrected formula 
of Ullah.$^6$ For illustration, we finally compute a simple three-vector 
integral, which can be evaluated with an independent method also. As it 
should be, we find the same answer with both methods.

\subsection{The two-vector integral}

Consider the arbitrary two-vector integral $\la M\ra= \la\vec m,\vec n\ra$, 
where the first column vector of $M$ is denoted by $\vec m$ and the second by 
$\vec n$. In this case, Eq.~(\ref{OR_res}) leads directly to the 
following expression:
\begin{equation}
\la\vec m,\vec n\ra = \left(\ts \frac{N-1}{2}\right)_{\bar n/2}^{-1}
\sum_{\vec\kappa} {\vec n \choose \vec\kappa} \; (-1)^{(\bar n-\bar\kappa)/2} 
\left\{ \prod_{i=1}^N \left(\ts\frac{1}{2}\right)_{\kappa_i/2} \right\}
\left(\ts\frac{1}{2}\right)_{(\bar n-\bar\kappa)/2} \; 
\la\vec m+\vec n-\vec\kappa\ra \; .
\label{O2_res}\end{equation}
The sum runs over such $\vec\kappa$ only, for which all components are 
even. A bar over a vector quantity denotes, as before, the sum of all its 
components. This formula has already been used in Ref.~$^{16}$ to calculate
various two-vector integrals. The numerical tests performed in parallel
confirm its validity.

For later purpose we use Eq.~(\ref{O2_res}) to evaluate the following
simple integral:
\begin{equation}
\lla\begin{pmatrix} 1 & 1\\ 1 & 1\\ 0 & 0\\ \vdots & \vdots
\end{pmatrix}\rra = \frac{2}{N-1} \; (-1) \; \frac{1}{2} \; \lla 
\begin{pmatrix} 2 \\ 2 \\ 0 \\ \vdots \end{pmatrix} \rra = 
\frac{-1}{(N-1)N(N+2)} \; .
\label{O2_example}\end{equation}
Note that the same result can be obtained by an indirect 
method$^{12}$ also. \\

In principle, an integration formula for general two-vector integrals has
already been published some time ago.$^6$ After the correction of two 
misprints, it reads:
\begin{equation}
\begin{split}
\overline{ \prod_i u_i^{2m_i} v_i^{2n_i}} &=
\pi^{-N+1} \; 2^{-2N+4-2\sum(m_i+n_i)} \;
\frac{\Gamma(N-1) \; \Gamma\left[N-1+\sum_i (m_i+n_i)\right]}
{\Gamma\left[\sum_i m_i +(N-1)/2\right] \;
\Gamma\left[\sum_i n_i +(N-1)/2\right]} \\
&\qquad\times
\sum_{k_1,\ldots,k_N,l_1,\ldots,l_N = 0,\ldots,0}%
^{2m_1,\ldots,2m_N,2n_1,\ldots,2n_N} (-1)^{\sum_i l_i} \\
&\qquad\times
\frac{\prod_i {2m_i\choose k_i}{2n_i\choose l_i} \;
\Gamma[(k_i+l_i+1)/2]\; \Gamma[m_i+n_i-(k_i+l_i-1)/2]}{
\Gamma[N/2+\sum_i (k_i+l_i)/2]\; 
\Gamma[N/2+\sum_i (m_i+n_i) - \sum_i (k_i+l_i)/2]} \; ,
\end{split}
\label{A_Ullah2v}\end{equation}
where $\forall i : k_i+l_i$ must be even. Here the original notation of 
Ref.~$^6$ is used. The corrections concern the first line, where the 
nominator has been multiplied with 
$\Gamma\left[N-1+\sum_i (m_i+n_i)\right]$, and the sum over 
$k_1,\ldots,k_N,l_1,\ldots,l_N$, where the vector-indices must start with 
zeros instead of ones. Finally the notation is quite unfortunate, as it 
seems to prohibit monomials with odd powers, though there is no reason 
for it. Indeed, Eq.~(\ref{A_Ullah2v}) holds in those cases 
as well. This can be checked, for instance, by computing the integral 
(\ref{O2_example}) with the help of Eq.~(\ref{A_Ullah2v}), setting 
$m_1=m_2=n_1=n_2=1/2$. Using the notation adopted in the present work,
Eq.~(\ref{A_Ullah2v}) reads:
\begin{equation}
\la\vec m,\vec n\ra = \frac{ (N-1)_{(\bar m+\bar n)/2}}{2^{\bar m+\bar n} \;
\left(\frac{N-1}{2}\right)_{\bar m/2} \;
\left(\frac{N-1}{2}\right)_{\bar n/2}} \;
\sum_{\vec k,\vec l} {\vec m \choose \vec k} \; 
{\vec n \choose \vec l} \; (-1)^{\bar l} \; 
\la\vec k+\vec l\ra \; \la\vec m-\vec k+\vec n-\vec l\ra \; .
\label{MUllah2v}\end{equation}
If we compare the integration formulas (\ref{O2_res}) and (\ref{MUllah2v}),
they differ considerably. It seems rather difficult to prove their 
equivalence directly. Note moreover, that our result is much simpler, 
because there the sum runs over a single vector-index only.

\subsection{A simple three-vector integral}

The three-vector integral considered here, is chosen because of its 
simplicity and because it may be evaluated using an indirect method, which 
allows to crosscheck the result. We will compute the integral $\la M\ra$ with 
\begin{equation}
M = \begin{pmatrix} 2 & 0 & 0\\
0 & 2 & 0\\
0 & 0 & 2\\
0 & 0 & 0\\
\vdots & \vdots & \vdots\end{pmatrix} \; .
\end{equation}
Henceforth we will skip those parts of the column vectors which are zero 
anyway. Using our recursion formula~(\ref{OR_res}) the three-vector 
integral $\la M\ra$ can be reduced to a linear combination of two-vector 
integrals, for which we already have a closed expression, namely
Eq.~(\ref{O2_res}). Thus, the evaluation of $\la M\ra$ needs only a 
few steps:
\begin{align}
\lla\begin{pmatrix} 2 & 0 & 0\\
0 & 2 & 0\\
0 & 0 & 2\end{pmatrix}\rra &= \frac{2}{N-2} \left\{ - \sum_K \left(\left. 
\begin{pmatrix} 0\\ 0\\ 2\end{pmatrix} \right| K\right)
\left(\frac{1}{2}\right)_{\bar k_1/2} 
\left(\frac{1}{2}\right)_{\bar k_2/2}
\lla\begin{pmatrix} 2 & 0\\ 0 & 2\\ 0 & 0\end{pmatrix} + K \rra +
\right. \notag\\
&\qquad + \left.
\frac{1}{2} \lla\begin{pmatrix}2 & 0\\ 0 & 2\end{pmatrix}\rra \right\} \notag\\
&= \frac{2}{N-2} \left\{ - \frac{1}{2} \left[ 
\lla\begin{pmatrix} 2 & 0\\ 0 & 2\\ 2 & 0\end{pmatrix}\rra +
\lla\begin{pmatrix} 2 & 0\\ 0 & 2\\ 0 & 2\end{pmatrix}\rra \right] + 
\frac{1}{2}\; \frac{N+1}{(N-1)N(N+2)} \right\} \notag\\
&= \frac{1}{N-2} \left\{ \frac{N+1}{(N-1)N(N+2)} - 2 \;
\frac{N+3}{(N-1)N(N+2)(N+4)} \right\} \notag\\
&= \frac{N^2+3N-2}{(N-2)(N-1)N(N+2)(N+4)} \; .
\label{R3_res1}\end{align}
The result is expressed as a rational function, where care has been 
taken, that nominator and denominator have no more common factors. \\

Alternatively we may compute $\la M\ra$, starting from the following 
identity:
\begin{equation}
\left(\sum_i w_{i1}^2\right)\left(\sum_j w_{j2}^2\right)
\left(\sum_k w_{k3}^2\right) = 1 \; ,
\label{A_id1}\end{equation}
which holds for an arbitrary $w\in O(N)$. Now we expand the products on 
the {\it l.h.s.} and integrate on both sides over the group. This gives:
\begin{equation}
N(N-1)(N-2) \lla\begin{pmatrix} 2 & 0 & 0\\
0 & 2 & 0\\
0 & 0 & 2\end{pmatrix}\rra + 3N(N-1) \lla\begin{pmatrix} 2 & 0\\
2 & 0\\
0 & 2\end{pmatrix}\rra + N \lla\begin{pmatrix} 2 \\ 2 \\ 2 \end{pmatrix}\rra 
= 1 \; .
\label{vecRex1}\end{equation}
It allows to express the three-vector integral $\la M\ra$ as a linear 
combination of a one-vector and a two-vector integral. According to the
Eqs.~(\ref{O1_res}) and (\ref{O2_res}), these integrals are given
by:
\begin{equation}
\lla\begin{pmatrix} 2\\2\\2\end{pmatrix}\rra = \frac{1}{N(N+2)(N+4)} \; , 
\quad \lla\begin{pmatrix} 2 & 0\\
2 & 0\\
0 & 2\end{pmatrix}\rra = \frac{N+3}{(N-1)N(N+2)(N+4)} \; .
\end{equation}
Thus we finally obtain:
\begin{align}
\lla \begin{pmatrix} 2 & 0 & 0\\
0 & 2 & 0\\
0 & 0 & 2\end{pmatrix}\rra &= 
\frac{(N+2)(N+4) -3(N+3)+1}{(N-2)(N-1)N(N+2)(N+4)} \notag\\
&= \frac{N^2+3N-2}{(N-2)(N-1)N(N+2)(N+4)} \; .
\end{align}
As expected, the result coincides with the one above, {\it i.e.} 
Eq.~(\ref{R3_res1}). Here the indirect method worked so well because we 
first wrote down the identities (\ref{A_id1}) and (\ref{vecRex1}), and then 
chose our particular example $\la M\ra$. However, if the value of a certain 
integral is needed, one would have to guess useful identities which allow to 
express the integral by a linear combination of simpler ones, a procedure wich 
is certainly very difficult. In contrast to that, the recursion 
formula~(\ref{OR_res}) always provides a well defined finite procedure, for
the computation of any integral.

\section{\label{S}Conclusions}

To summarize, we have derived a recursion formula, which expresses an 
arbitrary $R$-vector integral over the orthogonal group as a linear combination 
of $(R$$-$$1)$-vector integrals. It allows to successively evaluate the 
group integral of any finite monomial in the matrix elements of the
group. The simplicity of the result depends primarily on $R$, the number of 
column or row vectors involved, and only secondarily on the degree of the 
monomial in question. The result is always given as a finite sum of
partial fractions in $N$.

As an immediate consequence of the general result, we obtained a closed 
integration formula for arbitrary two-vector integrals, which is quite 
different and much simpler than the corrected, previously known result. 

In principle a similar recursion formula can be obtained for integrals over 
the unitary group also. To that end one should consider monomials 
in the real and imaginary parts of the matrix elements. Though the derivation
along the lines of the orthogonal case is rather straight forward, the 
resulting expressions are much more involved. It seem that the simple result 
for the case $R=1$ is only an exception. More work is clearly necessary 
to clarify the situation in this case.

\section*{Acknowledgments} 

I thank T. H. Seligman for many discussions, helpful comments, and critically 
reading the manuscript.

\section*{References}

\begin{tabbing}
$^10$ \= tralala \kill

$^1$ \> H. Weyl.
\newblock {\em The classical groups}. Princeton: Princeton U.P., 1939. \\

$^2$ \> E. Cartan.  \newblock {\em Abh. Math. Sem. Univ. Hamburg}, 11:116,
1935. \\

$^3$ \> L. K. Hua.
\newblock {\em Harmonic analysis of functions of several complex variables}. \\
\> \newblock Providence, RI: American Mathematical Society, 1963. \\

$^4$ \> M.~L. Mehta.
\newblock {\em Random Matrices and the statistical theory of energy levels}.\\
\> \newblock Academic Press, Boston, 1991. \\

$^5$ \> N. Ullah and C. E. Porter.
\newblock {\em Phys. Rev.}, 132(2):948, 1963. \\

$^6$ \> N. Ullah. \newblock {\em Nucl. Phys.}, 58:65, 1964. \\

$^7$ \> P.~A. Mello and T.~H. Seligman.
\newblock {\em Nucl. Phys. A}, 344:489, 1980. \\

$^8$ \> T.~Prosen, T.~H. Seligman, and H.~A.~Weidenm\" uller.
\newblock {\em Europhys. Lett.}, 55(1):12, 2001. \\

$^9$ \> T. H. Seligman. \newblock private communication. \\

$^{10}$ \> A.~Haar.  \newblock {\em Ann. Math.}, 34:147--169, 1933. \\

$^{11}$ \> M.~Abramowitz and I.~A. Stegun, editors.
\newblock {\em Handbook of mathematical functions}.\\
\> \newblock Dover, New York, 1964. \\

$^{12}$ \> T.~A. Brody, J. Flores, J. B. French, P. A. Mello, A. Pandey, and
S. S. M. Wong.\\
\> \newblock {\em Rev. Mod. Phys.}, 53(3):385, July 1981. \\

$^{13}$ \> F.~J. Dyson and M.~L. Metha.  
\newblock {\em J. Math. Phys.}, 4:701, 1963. \\

$^{14}$ \>  C.~E. Porter, editor.
\newblock {\em Statistical theories of spectra: Fluctuations}. New York:
  Academic Press,\\
\> 1965. \\

$^{15}$ \> F.-M. Dittes, I.~Rotter, and T.~H. Seligman.
\newblock {\em Phys. Lett. A}, 158:14, 1991. \\

$^{16}$ \> T. Gorin and T. H. Seligman.
\newblock {\em nlin.CD}/ 0101018, January 2001. \\

\end{tabbing}

\end{document}